\documentclass[]{aastex631}

\submitjournal{PSJ}

\shorttitle{BYORP and Dissipation in Binary Asteroids}
\shortauthors{\'Cuk et al.}

\graphicspath{{./}{figures/}}

\begin{document}

\title{BYORP and Dissipation in Binary Asteroids: Lessons from DART}

\correspondingauthor{Matija \'Cuk}
\email{mcuk@seti.org}

\author[0000-0003-1226-7960]{Matija \'Cuk}
\affiliation{SETI Institute, 189 N Bernardo Ave, Suite 200, Mountain View, CA 94043, USA} 

\author[0000-0002-3544-298X]{Harrison Agrusa}
\affiliation{Universit\'e C\^ote d'Azur, Observatoire de la C\^ote d'Azur, CNRS, Laboratoire Lagrange, Nice, France}

\author[0009-0000-2266-6266]{Rachel H. Cueva}
\affiliation{Smead Aerospace Engineering Sciences, University of Colorado Boulder, 3775 Discovery Dr., Boulder, CO 80303, USA}

\author[0000-0001-7537-4996]{Fabio Ferrari}
\affiliation{Department of Aerospace Science and Technology, Politecnico di Milano, Italy}

\author[0000-0002-1821-5689]{Masatoshi Hirabayashi}
\affiliation{Daniel Guggenheim School of Aerospace Engineering, Georgia Institute of Technology, Atlanta, Georgia 30332, USA}

\author[0000-0002-4952-9007]{Seth A. Jacobson}
\affiliation{Department of Earth and Environmental Sciences, Michigan State University, 288 Farm Ln, Rm 207, East Lansing, MI 48824, USA}

\author[0000-0002-1847-4795]{Jay McMahon}
\affiliation{Smead Aerospace Engineering Sciences, University of Colorado Boulder, 3775 Discovery Dr., Boulder, CO 80303, USA}

\author[0000-0002-0884-1993]{Patrick Michel}
\affiliation{Universit\'e C\^ote d'Azur, Obserbatoire de la C\^ote d'Azur, CNRS, Laboratoire Lagrange, Nice, France}

\author[0000-0003-3610-5480]{Paul S{\'a}nchez}
\affiliation{Colorado Center for Astrodynamics Research, University of Colorado Boulder, 3775 Discovery Dr, Boulder, CO 80303, USA}

\author[0000-0003-0558-3842]{Daniel J. Scheeres}
\affiliation{Smead Aerospace Engineering Sciences, University of Colorado Boulder, 3775 Discovery Dr, Boulder, CO 80303, USA}

\author[0000-0001-5475-9379]{Stephen Schwartz}
\affiliation{Planetary Science Institute, 1700 East Fort Lowell, Suite 106, Tucson, AZ 85719-2395, USA}
\affiliation{Instituto de Fisica Aplicada a las Ciencias y las Tecnologias, Universidad de Alicante, 03690 Sant Vicent del Raspeig, Spain}

\author[0000-0002-0906-1761]{Kevin J. Walsh}
\affiliation{Southwest Research Institute, 1050 Walnut St. Suite 400, Boulder, CO 80302, USA}

\author[0000-0003-4045-9046]{Yun Zhang}
\affiliation{Climate \& Space Sciences and Engineering, University of Michigan, Ann Arbor, MI 48109, USA }

\begin{abstract}
The Near-Earth binary asteroid Didymos was the target of a planetary defense demonstration mission DART in September 2022. The smaller binary component, Dimorphos, was impacted by the spacecraft in order to measure momentum transfer in kinetic impacts into rubble piles. DART and associated Earth-based observation campaigns have provided a wealth of scientific data on the Didymos-Dimorphos binary. DART revealed a largely oblate and ellipsoidal shape of Dimorphos before the impact, while the post-impact observations suggest that Dimorphos now has a prolate shape. Here we add those data points to the known properties of small binary asteroids and propose new paradigms of the radiative binary YORP (BYORP) effect as well as tidal dissipation in small binaries. We find that relatively spheroidal bodies like Dimorphos made of small debris may experience a weaker and more size-dependent BYORP effect than previously thought. This could explain the observed values of period drift in several well-characterized binaries. We also propose that energy dissipation in small binaries is dominated by relatively brief episodes of large-scale movement of (likely surface) materials, rather than long-term steady-state tidal dissipation. We propose that one such episode was triggered on Dimorphos by the DART impact. Depending on the longevity of this high-dissipation regime, it is possible that Dimorphos will be more dynamically relaxed in time for the Hera mission than it was in the weeks following the impact.     
\end{abstract}

\section{Introduction} \label{sec:intro}

Binaries are found among many dynamical classes of Solar System small bodies \citep{mar15, wal15, nol08}, with the binary systems greatly varying in structure, origin, and evolution. Here we discuss a subset termed ``small binary asteroids'' (SBAs), each consisting of a km-sized rapidly rotating primary and a relatively large (with a radius ratio $R_2/R_1>0.2$) synchronously rotating and elongated secondary \citep{wal15}. Such systems are sometimes termed ``singly-synchronous binaries'' to distinguish them from fully synchronous pairs where both components are spin-locked. 

Small binary asteroids are thought to be primarily formed by the spin-up of the primary to the critical rotation by the Yarkovsky-O'Keefe-Radzievskii-Paddack (YORP) effect \citep{rub00, bot06}, followed by the formation of the secondary which may be gradual \citep{wal08} or may happen in an episode of major instability \citep{jac11b,tardivel2018}. Small binaries may have somewhat similar origin mechanisms with heliocentric asteroid pairs \citep{vok08, pra10}, which also appear to originate through rotational disruption. 

As the typical SBA secondaries have irregular shapes and are predominantly in synchronous rotation state, the asymmetry between the re-emission of thermal radiation from their leading and trailing hemispheres (binary YORP, or BYORP, effect) should evolve their orbits on short ($<10^5$~yr for a typical NEA binary)  timescales \citep{cuk05, mcm10a}. Rival paradigms of small binary asteroids evolution have been proposed: that of fast-forming and fast-destroyed ephemeral systems ruled by radiation forces \citep{cuk07}, and that of largely long-lived stable systems in which tides and BYORP are in equilibrium \citep{jac11a}. Tide-BYORP equilibrium would not only require tidal dissipation in small rubble piles to be higher than previously estimated \citep[e.g.,][]{gol09}, but for smaller bodies to respond more strongly to changing gravitational forces  \citep{jac11a}. Larger deformation due to gravitational forces in smaller objects (measured by Love number $k_2$) is opposite from expectations from classical tidal theory \citep{md99}, and may indicate tidal response dominated by near-surface materials \citep{nim19}.

The first binary with observationally measured mutual orbital evolution, 1996~FG3, showed almost no orbital period change at all, with the observed semimajor axis migration rate of $\dot{a}=0.07 \pm 0.34$~cm~yr$^{-1}$ \citep{sch15}. This was widely seen as supporting evidence for tide-BYORP equilibrium. More recent observational results \citep{sch21} for two other asteroids, show one (2001~SL9) having a rapidly shrinking orbit with $\dot{a}=-2.8 \pm 0.2$~cm~yr$^{-1}$ or $\dot{a}=-5.1 \pm 0.2$~cm~yr$^{-1}$ (presumably due to BYORP), while the other (Moshup) has a slowly expanding orbit with $\dot{a}=1.2 \pm 0.3$~cm~yr$^{-1}$. It appears that not all small binaries are uniformly in equilibrium with a zero net orbital evolution rate. 

On September 26, 2022, the Double Asteroid Redirection Test (DART) spacecraft impacted Dimorphos, the secondary of the small binary asteroid Didymos. While the DART mission was conducted as a planetary defense test, it also offered us a first close-up look at a small binary asteroid \citep{riv21, che23, dal23}. One unexpected early result was that Dimorphos may be (or at least was before the impact) an oblate, rather than prolate, body \citep{Daly2024}. Interestingly, ground-based data implies that Didymos had a very slow secular drift in the mutual orbit period, a possible indication of its proximity to the tide-BYORP equilibrium \citep{sch22, pra22, nai22}. Some of the properties of post-impact mutual orbit, such as relatively low eccentricity and slow apsidal precession \citep{sch24} have been interpreted as being indicative of spin-orbit coupling including a {\it prolate} Dimorphos \citep{bor90, cuk10}. The rendezvous of the Hera spacecraft with Didymos in late 2026 should resolve some of the open questions about the current state of the Didymos system \citep{mic22}. 

In this paper, we will re-visit the idea of tide-BYORP equilibrium in small binary asteroids in light of some new theoretical and experimental results. The new theoretical development is the inclusion of the centrifugal force in the calculation of tidal dissipation, which can be very important for primaries that are spinning very close to the breakup limit. We will also consider the plausibility of the widespread tide-BYORP equilibrium in the context of the observed binary population. The experimental results come from the DART mission, as they suggest that the secondary is much closer to an ellipsoidal shape than single asteroids, with possible implications for the BYORP effect. Additionally, it appears that the Dimorphos has acquired a more energetically favored elongated shape following the perturbation from the DART impact, with implications for the nature of available dissipation mechanisms. We will evaluate how these ideas change our understanding of small binary asteroid dynamics and our interpretation of the properties of the small binary asteroid population. Our goal is to arrive at an interpretation of theory and data consistent with all available observations and to produce hypotheses to be tested by the Hera mission and other future observations.

\section{Binary YORP for Spheroidal Secondaries}\label{sec:byorp}

Early studies of radiation effects like YORP and BYORP required modelling a diverse array of asteroid shapes before many real asteroids had resolved shapes. The initial BYORP paper, \citet{cuk05}, followed the YORP modelling approach of \citet{vok02} and used random Gaussian spheroids to approximate asteroid shapes. In particular, the model of \citet{mui98} was used in which the coefficients of spherical harmonics are selected to be random, with certain constraints on their variation. The resulting artificial asteroids are decidedly ``lumpy'' \citep[see][for examples]{vok02}, but generally in agreement with the few asteroid shapes known precisely at the time, such as Toutatis and Eros, plus other small bodies such as Deimos. Later work on BYORP often focused on specific bodies \citep[e.g.,][]{mcm10b}, rather than the general population. 

When the orbital drift was first detected in a binary asteroid 1996~FG3 \citep{sch15}, it was much smaller than predicted by BYORP theories and actually consistent with zero. These early results gave impetus to the hypothesis of tide-BYORP equilibrium \citep{jac11a}, which posits that the bulk of observed asteroidal satellites are in balance between outward tidal migration and inward BYORP-driven evolution. For this equilibrium to be common across a range of small binary asteroid sizes, tidal response of the primary had to be {\it inversely} proportional to diameter, which is opposite the prediction of most classical \citep{md99} and some modern \citep{gol09} theories of tidal dissipation. Further detections of fast migration of 2001~SL9 and slow evolution of Moshup \citep{sch21}, as well as the near-zero drift of Didymos \citep{sch22} were interpreted in the context of the tide-BYORP equilibrium paradigm. For example, the slow evolution of Moshup (1999~KW9) was seen as a difference of opposing accelerations from tides and BYORP, which are not exactly in equilibrium. 

The imaging of Didymos and Dimorphos by the DART spacecraft just before its impact on Dimorphos provided the first close-up views of a small binary asteroid \citep{dal23}. While the look of a top-shaped primary is already familiar from radar imaging \citep[and was also seen in non-binary mission targets Bennu and Ryugu;][]{ost06, bus11, bro11, nol13, bec15, nai15b, bar19, wat19}, its shape was more oblate than radar observed \citep{nol13}, implying a new class of top-shaped objects. The ellipsoidal ``debris-pile'' look of Dimorphos was observed for the first time, whereby ``debris pile," we mean a rubble pile consisting only of elements much smaller than the whole at least on the surface. The secondary being made from smaller debris is not unexpected given some of the satellite formation models \citep[e.g.,][]{wal08}, in which the secondary is made from a stream of material lost from the primary. Note that we do not know whether there are large monolithic pieces inside of Dimorphos, but the surface consists only of blocks much smaller in size than the asteroid itself. Dimorphos's surface structure is thought to be weak and likely susceptible to mobilization \citep{rad24}. 

This leads to an interesting possibility that the general nature of the secondaries in small binary asteroids is that of ellipsoidal piles of relatively small-blocked debris. We do know that most secondaries are elongated from lightcurves \citep{pra16} and also from radar images of several objects \citep{ost06, bec15, nai15b}. Clearly, not all secondaries are smoothly ellipsoidal, as demonstrated by Lucy's mission flyby target Dinkinesh \citep{lev23}, whose secondary is made of two distinct, relatively angular components. Lightcurve data allow us to put some limit on how common systems like Dinkinesh are, with 2:1 aspect ratio for the satellite. \citet{pra16} found that secondary equatorial elongation appears to have an upper limit of about 1.5, implying that systems like Dinkinesh are uncommon (or even belong to a separate and previously unobserved binary type found only in the main belt). Selam's interesting dynamical history gives us invaluable information on binary physics, but we think that lightcurve statistics strongly indicate this is a relatively rare evolution path. Therefore, we will assume that the typical small binary's secondary is, in terms of block size and overall smoothness, more like Dimorphos than Dinkinesh's contact-binary moon. Dimorphos's own apparent oblate shape is not typical of SMA secondaries and will be addressed in the next Section. Prolate vs. oblate issue aside, we will assume that the majority of secondaries have a Dimorphos-like relatively smooth ellipsoidal appearance, and explore the implications. 

The important aspect of this kind of rubble pile would be that the characteristic size of rubble fragments is presumably independent of the secondary's size, but depends on their material properties. If this is true, the surface roughness of the secondary would be constant in terms of absolute measurement (in meters), as long as the size distribution of the material does not change. Therefore, the secondary's figure would become closer to a smooth ellipsoid the larger the body is. In the limit of infinitely large bodies and insignificant surface roughness, there would be no BYORP torque. Therefore, intuition suggests that larger secondaries should have a relatively weaker BYORP acceleration, as long as they are made of rubble blocks of the same absolute size. By ``relatively weaker BYORP acceleration'' we mean that the characteristic BYORP coefficient $B$ \citep[that depends on shape alone;][]{jac11a} is actually smaller for larger secondaries, before taking into account their larger mass-to-surface-area ratio. This approximation applies only to typical secondaries that are a small fraction (few percent) of the primary's mass, and are thought to be made of relatively small debris from the primary's surface. A secondary that comprises a significant fraction of primary's mass would need to be more representative of the primary's makeup and include some larger fragments, making our simple model inapplicable.

In order to quantify this model, we constructed a framework for generating virtual ellipsoidal asteroids with randomly uneven surfaces. We start from a simple tri-axial ellipsoid with axes $a > b > c$. By definition, we set $c=1$. We then generate a network of points on the surface that are used to determine the orientation of flat triangular facets. This network is based on an icosahedron, and has at least twenty facets. Depending on how small we want the facets, we assign our model the number $i$, which is how many times the original facets were subdivided into four more. So for $i=2$ we have $20 \times 4^2=320$ facets, and for $i=4$ we have $20 \times 4^4 = 5120$ facets. The points themselves are assigned a random vertical distance from the reference ellipsoid, with the probability being uniformly distributed between $-h < r-r_0 < h$, where $0 < h < 1$ is some fraction of the shortest axis $c$, and $r$ and $r_0$ are radius-vectors of a vertex and its projection on the surface of the reference ellipsoid. When generating a synthetic asteroid, apart from ellipsoidal axis ratios, the only relevant parameters are $i$, determining the number of facets, and $h$, which determines the scale of the ``relief''. 

\begin{figure*}
\plotone{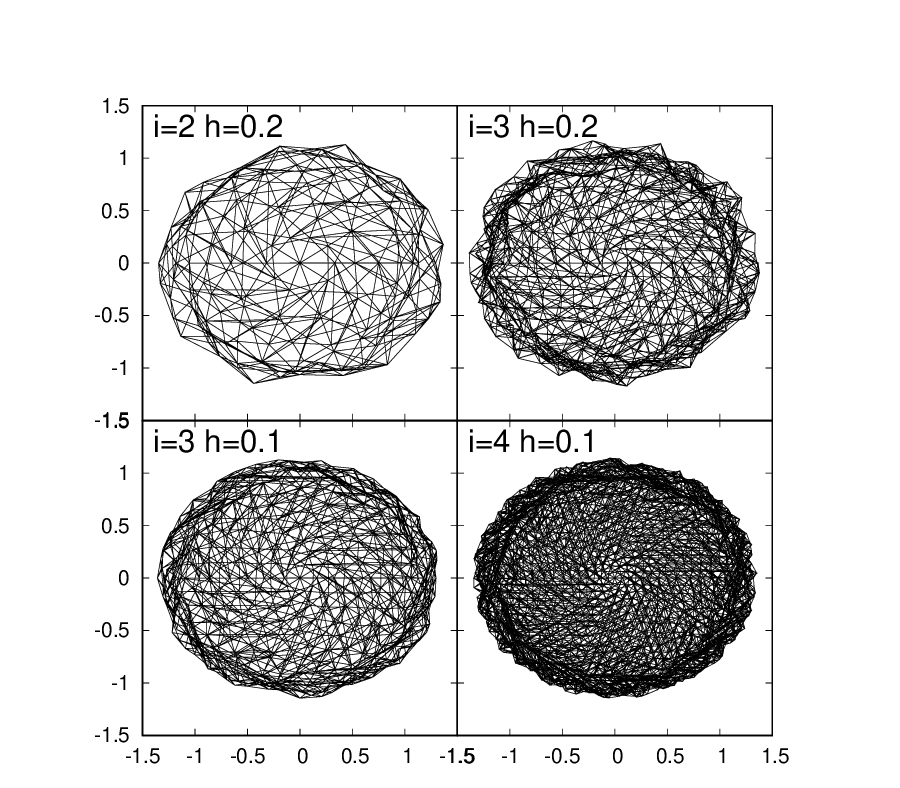}
\caption{Four different realizations of an ellipsoidal synthetic asteroid with axis ratios $a/c=1.3$ and $b/c=1.1$, with facet subdivision parameter $i$ and vertical relief parameter $h$ indicated in each panel (the scale is set by $c=1$). This illustration simply shows lines connecting vertices in numbering order, the asteroid being fully ``transparent''; the goal is simply to illustrate trends with changing $i$ and $h$.\label{shapes}}
\end{figure*}

On a real rubble-pile asteroid, parameters $d$, which determines the size of blocks on the surface, and $h$, which determines relief, are not independent and would be expected to be similar. As the area of the ellipsoid (for typical secondaries) is $ \approx 4 \pi (((ab)^{1.6}+(ac)^{1.6}+(bc)^{1.6})/3)^{1/1.6} \simeq 15-20$, the number of facets is $20 \times 4^i$ and the area of a triangular facet is $d^2 \sqrt{3}/4 \approx 0.433 a^2$ (where $d$ is a side of the facet), $d \simeq 1.6 2^{-4}$. So if we want to have $h \simeq d$, then we should use $i=4$ when $h=0.1$ and increase (or decrease) the number of the subdivisions of the icosahedron facets by one when $h$ shrinks (or increases) roughly by a factor of two, respectively. Figure \ref{shapes} shows some examples of the synthetic asteroids plotted as ``transparent'' bodies to illustrate how apparent roughness varies with changing $i$ and $h$.

After we have generated synthetic asteroid shapes, we can calculate the BYORP acceleration on them, following an approach based on \citet{cuk05}. We assume that the secondary is in synchronous rotation around the shortest axis of the reference ellipsoid, with the longest axis pointing to the primary. This assumes that the changes to the moments of inertia due to random positions of surface facets are negligible. While we could, in principle, determine principal moments of inertia after the synthetic shape was generated \citep[c.f.][]{cuk05}, that approach would necessarily assume uniform density, so it is not clear that it would be superior to our simpler method. We also assume that the plane of the binary's mutual orbit and the plane of its heliocentric orbit are the same, i.e., that the obliquity of the system is either zero or $180^{\circ}$. This is largely justified as small binary asteroids tend to have low (both prograde and retrograde) obliquities \citep{pra12}, likely due to the YORP evolution of the primary.  

When calculating the BYORP acceleration on the synthetic secondary, we assumed that all incoming energy from the Sun was absorbed by the surface of the facet (i.e., we ignore the reflected light), which is certainly not correct but not catastrophic given that even brighter asteroids typically have albedos of 20-25\%, and this cannot qualitatively change our results. We ignore any shadowing, meaning that the flux incoming on the facet depends only on the angle between the Sun and the facet normal (we set flux to zero if the Sun is more than $90^{\circ}$ from the facet normal). While shadowing (mostly by boulders) is admittedly present, it is expected to be less important for close-to ellipsoidal secondaries considered here than for bodies with more irregular (sometimes non-convex) shapes. We consider our model a first attempt at considering this new paradigm of secondary shape, and we think it is valuable to first consider the simplest model, which can be improved on later. Lastly, we ignore thermal inertia and assume that all energy is emitted by the facet immediately and perpendicular to the facet's surface. Prior work on the thermal inertia affecting the YORP effect found that the spin-changing torque is affected less than the obliquity-changing torque \citep{cap04}. Since BYORP is somewhat analogous to YORP, we do not expect thermal inertia to affect our conclusions about the dependence of the BYORP torque intensity on the constituent block size. Other work on BYORP also finds that leading coefficients governing BYORP are independent of thermal inertia \citep{mcm10a, mcm10b, cue24}. These assumptions could be tested with a more sophisticated model, such as \citet{ryo23} developed for the YORP effect.

For each realization of the shape of the secondary, we integrated absorbed and re-emitted flux over a large number of relative positions of the Sun and the secondary, uniformly distributed in longitude. By doing this, we are ignoring any effects from the heliocentric eccentricity of the binary. Our rationale was that outside of semisecular resonances with the Sun \citep{tou98, cuk10}, rapid precession of the satellite orbit ``smears out'' any effects of the heliocentric eccentricity. This is true for the planar case we are studying, but inclined binaries (with high-obliquity primaries) may experience variations in the radiation force that are periodic with the much slower precession of the primary's spin and the heliocentric orbit \citep{rub07}. We have also ignored shadowing by the primary; we note that the Yarkovsky-Schach effect, which is the principal consequence of shadowing, as well as effects driven by thermal radiation from the primary, should disappear for synchronous, zero-obliquity satellites \citep{mil87, rub06}. Still, the Yarkovsky-Schach effect is based on an assumption of a symmetrically shaped satellite, so it is not given that there would be no consequences for BYORP, which relies on an irregularly shaped secondary. We can only note that for our ellipsoidal secondaries, the main direction of the ``missing'' re-radiated energy during an eclipse would be radial and, therefore, would not change the semimajor axis. 

\begin{figure}[h!]
\vspace{-.15truein}
\hspace{0truein}
\begin{minipage}{2.8truein}
\centering
\includegraphics[scale=.8]{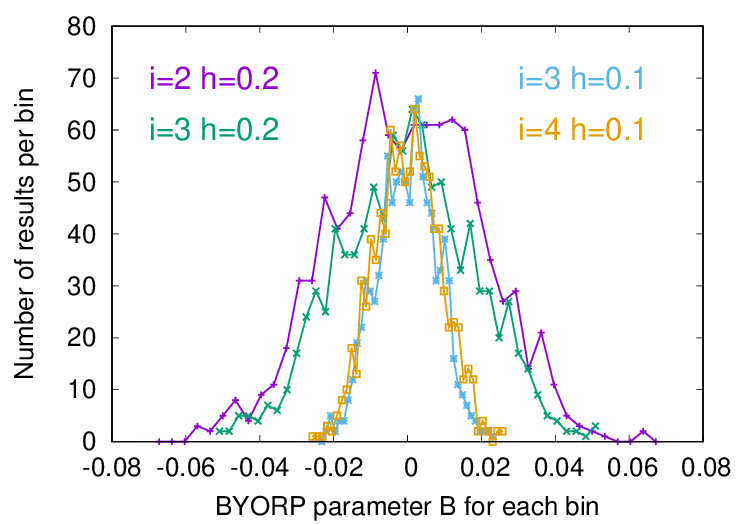}
\end{minipage}
\hspace{1.1 truein}
\vspace{-3.5in}
\begin{minipage}{2.9truein}
\centering
\vspace{-.1in}
\caption{Distribution of BYORP coefficients $B$ for sets of thousand synthetic asteroids for each of the four combinations of parameters $i$ and $h$ shown in Fig. \ref{shapes}. The BYORP force $F_B$ along the intermediate axis of the secondary is related to parameter $B$ by the relation $F_B=B R^2 \Phi/c$, where $R$ is the secondary's shortest radius, $\Phi$ is the solar flux experienced by the secondary, and $c$ is the speed of light. For each set of synthetic shapes, the largest absolute value of $B$ was used as the outer bound on the set of 40 equal-sized bins, and we plotted the number of synthetic shapes within each bin. As the sized bins vary from one set to the other, the plotted distributions cannot be directly quantitatively compared. For comparisons between dispersions for each distribution, see Fig. \ref{byorp}}\label{profile}
\end{minipage}
\vspace{3.4in}
\end{figure}

In order to quantify the BYORP effect, we follow \citet{jac11a} and use parameter $B$ which is basically a dimensionless effective cross section, and which needs to be multiplied by the square of the radius to be converted to dimensional units (further multiplication by solar radiation pressure is needed to obtain the BYORP force). Figure \ref{profile} plots the distribution of BYORP coefficients $B$ for one thousand realizations for each of the four combinations of parameters $i$ and $h$ illustrated in Fig.~\ref{shapes}. Note that we assumed that the secondary is in the principal axis, synchronous rotation, so this parameter $B$ is only the component of the BYORP recoil along the intermediate axis of the secondary, which can affect the mutual semimajor axis of the binary.  

It is clear from Fig.~\ref{profile} that a change in $h$ greatly changes the range of BYORP accelerations, while a change in $i$ does not have such a large effect. A larger range of $B$ BYORP coefficients stemming from larger vertical relief range parameter $h$ is intuitive, as a completely smooth ellipsoidal asteroid with $h=0$ should also have $B=0$. The lack of dependence on $i$, which determines the number of facets, is less obvious, and our interpretation is that the larger out-of-plane surface area available with more facets is counteracted by less stochastic behavior resulting from a larger number of smaller facets. In any case, we consider only those values of $i$ that give facet size comparable with vertical relief as valid, as we are trying to approximate a body made of polygonal blocks. If we go to higher values of $i$, the model acquires a ``hedgehog'' appearance, which is unrealistic, and would also be heavily affected by shadowing, which is not included. We conclude that the BYORP behavior of a synthetic asteroid depends mainly on $h$ and not $i$, although for other reasons, it is important to use a value of $i$ that is appropriate for that specific value of $h$.

To check that our calculations are correct, we also compute the radiation pressure force on the secondary, projected on the intermediate axis (like we did with BYORP). As we assume that the sunlight is incoming from an azimuthally symmetric source, we would expect there to be no net acceleration on the asteroid solely from incoming radiation for reasons of symmetry similar to the reason why there is no rotational torque from incoming radiation \citep{nes08}. Indeed, the radiation pressure equivalent of BYORP parameter $B$ is consistently in the $10^{-14}-10^{-13}$ range for the parameters used in Fig.~\ref{profile}, confirming that the way our shape models are handled does not cause any crude errors.

\begin{figure}[h!]
\vspace{-.15truein}
\hspace{0truein}
\begin{minipage}{2.8truein}
\centering
\includegraphics[scale=.8]{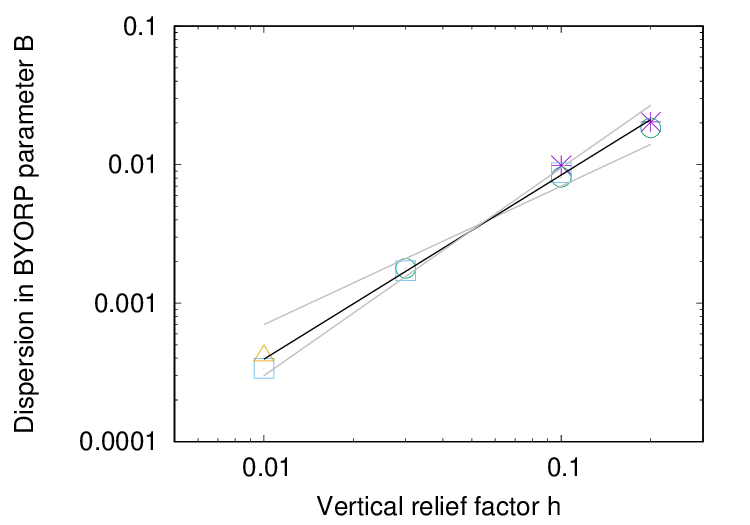}
\end{minipage}
\hspace{1.1 truein}
\vspace{-3.5in}
\begin{minipage}{2.5truein}
\centering
\vspace{-.1in}
\caption{Distribution of dispersions in BYORP parameters $B$ for a number of sets of synthetic asteroids, plotted against the shape parameter $h$. Different symbols indicate which value of $i$ (determining the number of facets) was used for the set: asterisks signify $i$=2, circles $i=3$, squares $i=4$ and the triangle indicates $i=5$. All sets included 1000 synthetic asteroids, except the $i=5$ one that had one hundred. Dispersion was calculated as $\sigma^2_B = {1 \over n} \sum^n_{i=0} (B_i-0)^2$. Solid lines plot approximating power laws with exponents 4/3 (black line), one (shallow grey line) and 3/2 (steeper grey line).}\label{byorp}
\end{minipage}
\vspace{3.4in}
\end{figure}

As the way our results are binned in Fig. \ref{profile} depends on the distributions' outliers, it is not fully correct to directly compare the plotted lines, so we have plotted the average dispersion of $B$ around zero for each set of synthetic asteroids (Fig. \ref{byorp}). It is clear from Fig. \ref{byorp} that the dependence of BYORP parameter $B$ on the number of facets (dictated by $i$) is very limited, but there is a very clear trend with surface relief quantified by $h$. If we assume a power-law dependence of $B$ on $h$, the best simple-fraction power exponent is 4/3, meaning that $B \propto h^{1.33}$. Note that this dependence of $B$ in our shape model is independent of (and adds to) the fundamental dependence of the BYORP timescale on the size of the components $t_B \propto R_1 R_2$ \citep{cuk05}, where $R_1$ and $R_2$ are the radii of the primary and the secondary.

\begin{figure}[h!]
\vspace{-.15truein}
\hspace{0truein}
\begin{minipage}{2.8truein}
\centering
\includegraphics[scale=.8]{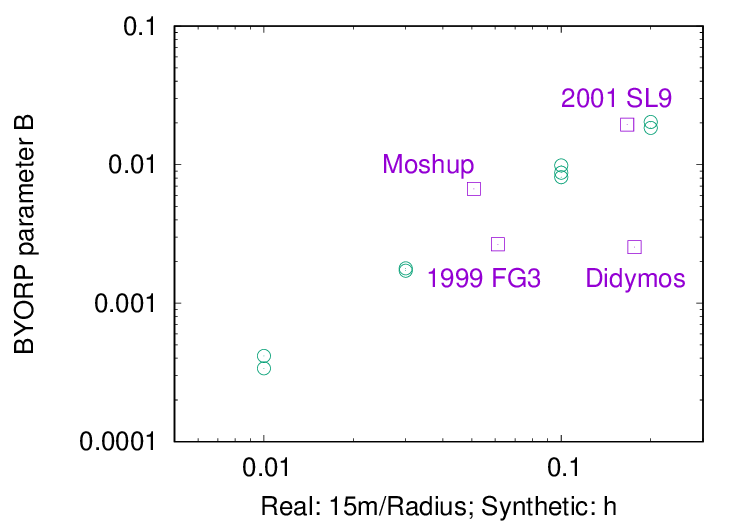}
\end{minipage}
\hspace{1.1 truein}
\vspace{-3.5in}
\begin{minipage}{2.5truein}
\centering
\vspace{-.1in}
\caption{Comparison of BYORP dispersions as cyan empty circles shown in Fig. \ref{byorp} with the BYORP parameters of real binary asteroid with detected mutual period drift \citep{sch15, sch21, sch22}. In order to assign the vertical roughness parameter $h$ to real secondaries, we divided 15~m by the mean radii of the secondaries. This 15-meter absolute roughness scale was chosen in order for the model to match the observed data, although the sizes of the largest boulders on Dimorphos are roughly consistent with this value \citep{paj24}. We ignored any tidal contribution to the orbital evolution when calculating the real asteroids' $B$ parameters.}\label{detected}
\end{minipage}
\vspace{3.4in}
\end{figure}

How does our model compare to the observed migration rate of real binary asteroids? Figure \ref{detected} shows the results for $B$ derived from our synthetic asteroids to values of $B$ derived from observations of real binaries. Here we assumed the slower of the two potential values for 2001~SL9 and used error bars to represent the $B$ value for 1999~FG3. In making this plot, we ignored any contribution from tides to the mutual orbital evolution of the observed pairs, and attributed all of the secondary's migration to BYORP. A source of greater uncertainty is the relationship we had to assume between the vertical relief parameter $h$ and the physical radii of real bodies. Assuming that the size of the surface irregularity is due to the blocks of rubble, and assuming that the size of these blocks is independent of asteroid size, then $h$ and the body's size should be inversely proportional to each other. This is a direct consequence of $h$ being defined as a fraction of the secondary's radius, while the roughness it is measuring is presumably independent of the size of the entire body. We somewhat arbitrarily set the conversion $h = 15~{\rm m}/R$ for observed asteroids, and ignored the likely differences in block size between different compositional classes.

Figure \ref{detected} implies that the relatively slow migration rate of Moshup \citep{sch21} and the near-zero migration rate of 1999~FG3 should be considered at least partly due to the relatively large physical size of the secondaries in those systems, at least when compared to the much smaller (and faster migrating) secondary of 2001~SL9. Of course, this relies on the assumption that these secondaries are ellipsoidal in shape, and that the shape irregularities are dominated by relatively small blocks on the surface, as is the case for Dimorphos. The secondary of Moshup has the best-characterized shape in the sample \citep{ost06, sch06}, and while it is roughly ellipsoidal, its low-resolution model predicts a larger BYORP acceleration than observed \citep[][with a caveat that we are ignoring tides here] {mcm10b}. Note that the distribution of $B$ parameters centered on zero (Fig. \ref{profile}) is a consequence of our model being based on a symmetric ellipsoid. Shape models with larger asymmetries (including large craters or boulders) have non-zero centers of $B$ distributions \citep[cf. ][]{cue24}.

Based on Fig. \ref{detected}, it appears that (pre-DART) Didymos is actually the outlier in having very slow orbital migration compared to predictions. There are multiple possibilities that can explain this result. One is simply that Dimorphos had a $B$ parameter close to the center of the normal-like distribution in Fig. \ref{profile}. The predictions for $B$ plotted in Figs. \ref{byorp} and \ref{detected} are actually 1-sigma values for the relevant distributions, so having a value that is only 0.2 $\sigma$ from zero would have a probability of over 15\%. Given extreme uncertainties of all parameters involved we do not think that probabilities can be determined any more accurately, but it is clear that some asteroids will have very small BYORP parameters $B$ just due to chance.

Another possibility is that of BYORP-tide equilibrium \citep{jac11a} for pre-DART Didymos. We will address this possibility later in the paper as part of the wider reappraisal of dissipation within binaries, but we will just note here that the other three asteroids with detected BYORP do not seem to be in tide-BYORP equilibrium. The remaining possibility is that Didymos was not in synchronous rotation before the DART impact. We will explore that possibility in the following section.

\section{Dissipation in Dimorphos After the DART Impact}

Didymos system has arguably been better characterized after the DART impact than before, at least in terms of secondary's lightcurve which was not detected pre-impact. Lightcurve and other observations have enabled measurements of orbital precession \citep{pra24, sch24, nai24}. While an oblate shape of the components leads to prograde precession of the pericenter of the orbit, a prolate shape of a synchronously rotating component would lead to a retrograde contribution to the orbital precession \citep{bor90, cuk10}. Fitting of Dimorphos's post-DART orbit indicates that there is a substantial contribution to the orbital precession from the figure of Dimorphos \citep{pra24, sch24}, requiring some degree of alignment of the secondary's long axis with the line between components. \citet{pra24} interpret the post-DART lightcurve data as implying moderate librations of Dimorphos around synchronous state, and find that an axis ratio of $a/b=1.1-1.4$ for Dimorphos is consistent with the observations. 

It is well known that the equilibrium shape for a close-in synchronously rotating satellite is a prolate ellipsoid, due to the tidal field of the primary \citep[e.g.][]{kea14}. Therefore, it is possible that reshaping of Dimorphos from oblate to prolate shape would result in a lower energy state. This is only one of the energy changes in the system, the most notable being semimajor axis change due to the DART impact, but there are other effects that need to be taken into account, such as the rotational energy of Dimorphos and its own self-gravitating energy. In this section, we will try to estimate the relative magnitudes of these changes in energy and use them to make conclusions about short-term dissipation in the Didymos system. 

The energy of an elliptical orbit is $E_{orb}= - (G m_1 m_2)/(2a_o)$ \citep{md99}, where $G$ is the gravitational constant, $m_1$ and $m_2$ masses of the primary and secondary, and $a_o$ the mutual semimajor axis. We will use the absolute value of the current (post-DART) orbital energy of the Didymos system as the normalization factor for all other energy components (normalization will be denoted by a hat). As the orbital period changed from 11.92~h \citep{pra22, nai22} to 11.37~h \citep{tho23, sch24, nai24}, the change in the orbital energy is (using Kepler's third law that $a_o^3 \propto P^2$, where $P$ is orbital period):
\begin{equation}
\Delta \hat{E}_{orb} = - (P_f/P_i)^{2/3} + 1 = - 0.032
\label{delta_e_orb}
\end{equation}
Note that this is only about 20\% of the kinetic energy of the DART impact (i.e. $\hat{E}_{DART} = 0.16$), with most of the total energy carried away by fast ejecta \citep[$<10^4$~kg moving at $1.6\pm0.2$~km~s$^{-1}$,][]{ofe24}. Because the impact made the binary tighter, DART actually net-drained energy from the Didymos system (as the fast ejecta escaped rapidly). Slow ejecta was estimated by \citet{ofe24} to amount to about $10^6$~kg moving at $10$~m~s$^{-1}$, while \citet{kim23} estimate a mass of $>10^7$~kg, but moving at speeds $<1$~m/s. While the slow ejecta may have contributed significantly to momentum change of Dimorphos, it carried a negligible fraction of the impact's energy. 

Another change in the system's energy is associated with the reshaping of Dimorphos.  Here we will ignore mass loss (due to slow ejecta) as it is a relatively small fraction compared to Dimorphos's mass, causing smaller effects than considered in this work. We start by estimating the minimum self-gravitating energy of Dimorphos in isolation, assuming a volume-equivalent sphere with a diameter of $2 r_v=151$~m \citep{dal23}:
\begin{equation}
E_{sphere}= {3 \over 5}{G m_2^2 \over r_v}
\end{equation}
Comparing this energy to the gravitational energy of the system, we find $\hat{E}_{sphere}=-0.14$. In general, calculating the binding energy of triaxial ellipsoids is non-trivial \citep{dob18}. We can simplify our task by approximating the pre- and post-DART figures of Dimorphos as symmetric oblate and prolate ellipsoids, respectively. Pre-impact Dimorphos had main axes of $177\pm2$~m, $174\pm4$~m and $116\pm2$~m \citep{dal23}, so we adopt axes of $2a=2b=176$~m and $2c=116$~m. The expression for the binding energy of a uniform density oblate body is \citep{dan62, dob18}:
\begin{equation}
E_{pre}=E_{sphere} {r'_v \arccos(c/a) \over \sqrt{a^2-c^2}}
\label{oblate}
\end{equation}
where $r'_v=(a^2c)^{1/3}$ is self-consistently defined volumetric radius. Using the above-stated values for different diameters, we find that $E_{pre}=0.985 E_{sphere}$ (since both energies are negative, a sphere is a lower energy state). Post-impact, we will assume that Dimorphos has the same volume, and is now a prolate ellipsoid with axis ratios $a=1.3 b$ and $c=b$ \citep[a value in the middle of region allowed by ][]{pra24}. The expression for self-gravitating energy of a prolate ellipsoid is \citep{dan62, dob18}:
\begin{equation}
E_{post}=E_{sphere} {r'_v \ln(a/c + \sqrt{a^2/c^2-1}) \over \sqrt{a^2-c^2}}
\label{prolate}
\end{equation}
with $r'_v=(ac^2)^{1/3}$ for a prolate body. Using the axis ratios given above, we find that $E_{post}=0.994 E_{sphere}$, indicating that a prolate post-impact shape represents a lower energy state. Normalized to the orbital energy, the change in self-gravitational energy of Dimorphos is $\Delta \hat{E}_{G} = - 1.3 \times 10^{-3}$. 

While \citet{pra24} did not estimate the new polar radius, \citet{nai24} found, with some modeling assumptions, that the new semiaxes of Dimorphos are $a=96$~m, $b=74$~m, and $c=59$~m. This would make Dimorphos significantly tri-axial and therefore unsuitable to the analytical estimates of binding energy \citep{dob18}. In order to roughly estimate the energetics of the \citet{nai24} shape model, we made two estimates of its binding energy, one approximating it as a prolate, and one as an oblate body. The prolate model uses Eq. \ref{prolate} with $a=96$~m and $c'=\sqrt{bc}=66.1$~m, and gives us $E_{N1}=0.988 E_{sphere}$. The oblate model uses Eq. \ref{oblate}, $a'=\sqrt{ab}=84.3$~m and $c=59$~m, giving us $E_{N2}=0.989~E_{sphere}$. Therefore, \citet{nai24} model is very likely also has a lower self-gravitating energy than the pre-impact figure. We note that the reason for the higher energy state of the pre-impact Dimorphos is not it being oblate {\it per se}, but that its flattening was relatively large with $a/c=b/c > 1.5$.

However, the change in Dimorphos's shape also leads to two other effects that change the system's energy; one of them being the interaction with the gravity field of Didymos. The potential energy of a body in a gravity field of a point mass is \citep{tou94}:
\begin{equation}
E_{int}= - {G m_1 \over 2 R^3} \mathrm{tr}(I) + {3 \over 2}{G m_1 \over R^5} \mathbf{R} \cdot I \mathbf{R} 
\end{equation}
Where $\mathbf{R}$ is a vector between the centers of mass of the primary and the secondary, $R=|\mathbf{R}|$, and $I$ is the inertia matrix. If we use the principal moments of inertia, the matrix $I$ is diagonal, made up of the three principal moments of inertia $A$, $B$ and $C$ ($C\geq B\geq A$). If we assume that the secondary is in synchronous rotation, on a circular orbit, with the longest axis aligned with the primary, we get:
\begin{equation}
E_{int}= - {G m_1 \over 2 a_o^3} (B+C-2A)
\end{equation}
For a uniform-density ellipsoid, the principal moments of inertia are $A=m_2 (b^2+c^2)/5$,  $B=m_2 (c^2+a^2)/5$ and $C=m_2 (a^2+b^2)/5$. This makes the interaction energy:
\begin{equation}
E_{int}=-{G m_1 m_2 \over 10 a_o^3} (2a^2-b^2-c^2) = - |E_{orb}|{2a^2-b^2-c^2 \over 5a_o^2} 
\label{int2}
\end{equation}
Using the pre- and post-DART dimensions of Dimorphos, and taking into account the shrinking of the semimajor axis, we get $\hat{E}_{int}^{pre}=-6.1 \times 10^{-4}$ and $\hat{E}_{int}^{post}=-9.7 \times 10^{-4}$, so $\Delta \hat{E}_{int}= - 3.6 \times 10^{-4}$. Note that the difference between the binding energies of present Dimorphos (i.e. our assumed prolate spheroid with $a=1.3b$) and a sphere is slightly smaller than the former's interaction energy with the tidal field of Didymos, which may explain why the present shape is maintained. If we use the shape model of \citep{nai24}, we get $\hat{E}^{N}_{int}=-1.3 \times 10^{-3}$, as their shape model is overall more elongated. Interaction energy for the \citet{nai24} nominal shape is close to, but not larger than, the difference between the self-gravity of this ellipsoid and the sphere, but the uncertainties of this shape model allow for solutions that are energetically favored over a sphere. 

In addition to the interaction with the gravity field of the primary, a synchronous secondary also has kinetic energy of rotation (assuming rotation around the shortest axis):
\begin{equation}
E_{rot}= {1 \over 2} C n^2 = {1 \over 10} m_2 (a^2+b^2) {G (m_1+ m_2) \over a_o^3} \approx | E_{orb} | {a^2 + b^2 \over 5a_o^2}    
\end{equation}
Where the last step assumes $m_1 >> m_2$. We find that $\hat{E}_{rot}^{pre} = 2.13 \times 10^{-3}$ and $\hat{E}_{rot}^{post} = 1.88 \times 10^{-3}$, so $\Delta \hat{E}_{rot} = - 2.5 \times 10^{-4}$. However, not all of this energy is actually dissipated, as the angular momentum conservation requires the orbit to grow slightly to offset the deficit of spin angular momentum. For spin-orbit exchanges close to synchronous state, most of the energy actually goes into orbit and only a small fraction is lost \citep[][Section 4.9]{md99}. Therefore we will ignore the rotational energy changes (due to reshaping of Dimorphos and orbital period change) in our budget on energy dissipation within Dimorphos.

\begin{deluxetable*}{lrr}
\tablenum{1}
\vspace{-.2in}
\tablecaption{Estimates of energy changes in the Didymos system, normalized to the binary's total post-impact orbital energy of $1.38 \times 10^8$~J \citep[based on][]{dal23}. The first column assumes prolate post-impact Dimorphos with $a/c=1.3$, while the second column assumes the triaxial ellipsoidal model of \citet{nai24}. \label{table}}
\tablewidth{0pt}
\tablehead{\colhead{Term} & \colhead{$\Delta E$ (prolate)} & \colhead{$\Delta E$ \citep{nai24}} }
\startdata 
Energy of the DART impact &0.16\phantom{000} & 0.16\phantom{000} \\
Immediate mutual orbital energy change & - 0.032\phantom{00} & - 0.032\phantom{00}\\
Dimorphos binding energy change & - 0.0013\phantom{0} & $\approx$ - 0.0005\phantom{0}\\
Didymos-Dimorphos tidal interaction change & - 0.00036 & - 0.00073\\
Long-term orbital energy change &$\approx$ - 0.001\phantom{00} & $\approx$ - 0.001\phantom{00}\\
\enddata
\end{deluxetable*}

The above calculations demonstrate that pre-impact Didymos system was in a metastable energetic state that was relaxed during the impact. Putting aside the orbital energy for now, Dimorphos has acquired negative binding energy amounting to $\Delta \hat{E}_G = -3.7 \times 10^{-3}$, and also about an order of magnitude smaller negative energy of gravitational interaction with Didymos. These adjustments could not have been produced solely by impact and removal of ejecta, but required Dimorphos to adjust to the new orbital state, likely over a period of many orbits. Also, this adjustment involved large-scale motion of at least some of Dimorphos's building blocks relative to each other.

The large-scale changes in Dimorphos's shape through relaxation are not completely unexpected, given its appearance as a rubble pile. But it appears almost certain that Dimorphos was not relaxed before the impact, as its highly oblate body was energetically disfavored by self-gravitational energy, interactions with the tidal field of Didymos, and rotational energy (assuming synchronous rotation).
Therefore, there are clearly different regimes of Dimorphos's response to outside forces: a largely solid-body behavior before the impact (a ``granular solid'') and an aggregate response after the impact (``granular liquid'') \citep{murdoch-ast4-2015}. We can only speculate that the ``solid'' behavior may be due to surface forces such as Van der Waals \citep{san14}, friction \citep{zha17, rad19}, geometrical cohesion \citep{roz14, bar19}, which were overcome due to the DART impact. 

How long does the liquid behavior last, before the body is ``solid'' again? The fact that the shape and rotation of Dimorphos appear to have conformed to the time-averaged tidal potential of Didymos suggests that the relaxation was not instantaneous and must have lasted for at least several mutual orbits. Lightcurve monitoring after the impact \citep{sch24, pra24} seems to indicate changes in eccentricity, but these may be due to spin-orbit coupling rather than dissipation. On the other hand, reported decay of orbital period over one month \citep{gud23, sch24} could be a consequence of dissipation. One possible mechanism for orbital decay would be damping of eccentricity. Decay of semimajor axis of $\Delta a_o/a_o = 10^{-3} $ (equivalent to one-minute change in the orbital period) would require damping of $e=0.03$ to zero, or alternatively an eccentricity change from $e=0.044$ to $=0.03$. Since we are talking about $\Delta \hat{E}_e = - 10^{-3}$, this amount of energy damping would be significantly less than that involved in the reshaping of Dimorphos. Note that the current energy of gravitational interaction between Dimorphos and Didymos is of the same order (Eq. \ref{int2}), so there is a possibility that at least some of the period change is due to spin-orbit interactions. However, the fact that Dimorphos damped a larger amount of energy during reshaping, and settled back into close to synchronous (if somewhat excited) rotation does suggest large-scale dissipation, so this orbital period change being from eccentricity damping is a strong possibility. Since eccentricity forces the excited rotation states of elongated secondaries \citep{wis84, cuk10}, damping of eccentricity would be a natural continuation of reshaping and synchronous re-locking after the impact.   

Apart from eccentricity damping, interaction with ejecta may also play a role in the slow prolonged decay of Dimorphos's semimajor axis. In order for ejecta interaction to dominate the semimajor axis decay, on the order of 0.1\% of Dimorphos's mass would need to be on bound orbits and receive a kick on the order of its orbital velocity ($\simeq 0.1-0.2$~m/s) in order to explain this change in Dimorphos's orbit. This would require at least $10^7$~kg of debris to be ejected in a $0.08-0.24$~m/s range so as not to fall back onto Dimorphos right away or be ejected outright. This estimated mass would also need to be in bodies large enough not to be ejected by radiation pressure but interactions with Dimorphos. More modeling is needed to address this possibility, but we prefer the eccentricity damping explanation as the semimajor axis change roughly corresponds to the energy needed to dissipate the eccentricity expected from the impact.

The hypothesis of ongoing large-scale dissipation within Dimorphos through landslides after the impact will be tested by the Hera spacecraft. We can probably expect that the rotation of Dimorphos will be more regular and less excited than found by \citet{pra24} and \citet{sch24} by the time of the Hera mission. If Dimorphos has not yet fully settled into a rigid rubble pile regime, tidal and inertial forces on the surface of Dimorphos arising from excited rotation may trigger new landslides \citep{agr22}. Additionally, \citet{cue24} show that the classical tidal dissipation is not strong enough to damp dynamical excitation of the Didymos system between the epochs of DART and Hera missions. Therefore, if there are truly two distinct regimes of physical behavior among Dimorphos-like ellipsoidal secondaries (and this might apply to rubble piles in general), Hera observations will be very valuable in determining how the putative dynamical ``freezing'' happens, after which the body behaves more or less like a rigid object. 

\section{Implications for Binary Population}\label{sec:dis}

The two new ideas presented in this paper: 1) relatively weak and size-dependent BYORP effect for ellipsoidal secondaries, and 2) rubble piles having two dissipation regimes, an almost-rigid one and one experiencing highly dissipative landslides. The first idea is directly suggested by the appearance of Dimorphos before the DART impact, while the second is based on an apparent large-scale reshaping of and damping within Dimorphos after the collision. How do these ideas agree with the information we have on other binary systems?

We know of at least one binary system in which the secondary is well known to be in chaotic rotation, 1991~VH \citep{nai15a}. Additionally, outer satellites in several triple systems appear to be in non-synchronous rotation \citep{bro11, bec15}, as do some very small secondaries on eccentric orbits \citep{pra16}. It appears that larger separation and smaller size of secondaries are often associated with non-synchronous rotation, which would make sense from conventional tidal theory. Here we assume that the non-synchronous rotation is driven by eccentricity \citep{wis84, cuk10} rather than being a primitive state, so the relevant dissipation timescale for synchronizing the orbit is that of eccentricity damping, rather than direct rotational spindown \citep[the latter should not be dependent on the secondary's size, only distance][]{md99}. Smaller secondaries, being more often non-synchronous, are contrary to the idea that tidal response in rubble piles increases with decreasing size \citep{jac11a, nim19}, but more work is needed to confirm this. YORP spin-up could break synchronous lock for some smaller secondaries, but may not be effective in systems experiencing eccentricity-driven chaotic rotation of the secondary. Despite all these uncertainties, we argue that the existence of long-lasting excited dynamical states suggests that the secondaries are behaving more like rigid solids than easily deformable bodies, and implies that many secondaries may be in the ``frozen'' rubble pile state.  

The only other small binary (serendipitously) visited by a spacecraft is Dinkinesh with its moon Selam \citep{lev23}. Selam is a contact binary, made from two spheroidal but decidedly blocky components. It is natural to assume that Selam may be a product of a soft merger between two satellites, likely with some involvement of the BYORP effect \citep{cuk05}, as tidal orbital convergence is not expected for approximately equal-sized bodies. While Selam appears to be in synchronous rotation, its shape is most definitely not relaxed. Therefore we must conclude that the conditions that made Dimorphos``fluidize'' and reshape were not present during the formation and evolution of Selam. One possibility is that the low impact velocities that could have been experienced during Selam's presumed merger were insufficient to fluidize the entirety of the two lobes away from the points of contact. Note that the gravitational binding energy of Dimorphos and the energy of the DART impact are comparable, just like the kinetic energy of Selam's merger must have been comparable to its binding energy. Therefore, the total kinetic energy of the impact may be less important for starting landslides than the velocities involved. Collisional velocities generally tend to be lower around less massive primaries, so small binaries may harbor more soft-merger satellites than larger systems \citep[cf. ][]{agr24}

If the idea that most of the dissipation in binary asteroids happens during relatively short periods of instability is true, long-term tidal dissipation rates cannot be derived from the age of the system. \citet{pou24} consider binary asteroids that are members of heliocentric asteroid pairs, and try to estimate spin-synchronization and eccentricity-damping timescale for those binaries on the basis of the age of the heliocentric pair. They find that smaller binary asteroids appear to have smaller tidal parameters $Q/k_2$ ($Q$ being tidal quality factor and $k_2$ the tidal Love number), contrary to older tidal theories \citep{md99, gol09}, but in agreement with the models assuming dissipation close to the surface \citep{nim19}. Our interpretation is that smaller asteroid pairs form more often \citep[likely through YORP breakup][]{pra10}, and drift apart faster (due to the Yarkovsky effect), so the age of observed heliocentric pairs will be correlated with component size. If most of the dissipation in the binary within the pair happens during a short ``landslide'' phase following the pair formation (or later binary formation, if these are separate events), the ages of the pairs have no bearing on the long-term tidal properties of these bodies. 

\begin{figure}[h!]
\vspace{-.15truein}
\hspace{0truein}
\begin{minipage}{2.8truein}
\centering
\includegraphics[scale=.8]{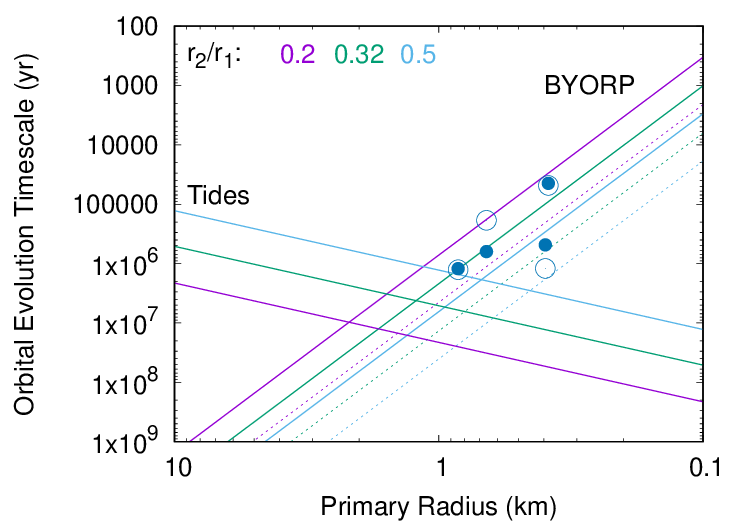}
\end{minipage}
\hspace{1.1 truein}
\vspace{-3.5in}
\begin{minipage}{3truein}
\centering
\vspace{-.1in}
\caption{Timescales for tidal evolution \citep[based on the model of ][]{gol09} and BYORP (this work) for different-sized primaries with synchronous satellites orbiting at a distance of 3 primary radii. Magenta lines are for systems with $r_2/r_1=0.2$, green lines for $r_2/r_1=0.32$ and blue lines for $r_2/r_1=0.5$. For BYORP timescales, solid lines plot values for the average heliocentric distance of 1~AU, while dashed lines plot values for 2.5~AU (the former are typical of NEAs while the latter apply to the inner main asteroid belt). The open circles plot the observed orbital evolution rates for select systems (cf Fig. \ref{detected}), while the filled circles plot the same values scaled to average heliocentric distance of 1~AU. Left to right, the four systems are 1996~FG3 ($r_2/r_1=0.29$), Moshup ($r_2/r_1=0.42$), Didymos ($r_2/r_1=0.2$) and 2001~SL9 ($r_2/r_1=0.24$).}\label{timescales}
\end{minipage}
\vspace{3.4in}
\end{figure}

Figure \ref{timescales} illustrates evolution timescales in our model of the BYORP effect for SBAs with spheroidal secondaries, compared with the tidal evolution of rubble piles using the \citet{gol09} model. The lifetimes are plotted as a function of primary size, with different color lines plotting the values for different secondary-to-primary size ratios. Note that this plot assumes a relatively close mutual separation of 3 primary radii, and that tidal dissipation weakens very fast with distance, while BYORP becomes slightly stronger. It is clear that the trends of the two processes with primary and secondary size are opposite, with BYORP, a radiation effect, being favored for small bodies. Tidal response is thought to become stronger with the body's size \citep{gol09}, but some theories predict an opposite dependence \citep{jac11b, nim19}. Smaller secondaries ($r_2/r_1 =0.2$) are expected to be dominated by BYORP for primary radii up to 2~km, while equal-size pairs are predicted to be tide-dominated for $r>300$~m at 1~AU and $r>200$~m at 2.5~AU.     

Our model of BYORP for ellipsoidal ``debris pile'' secondaries, combined with the \citet{gol09} tidal model, predicts that the observed orbital period drifts in small binary asteroids are primarily caused by BYORP and not tides. Given the parameters of the system, the observed evolution of 1996~FG3 should be closest to having a significant tide-driven component. Our model also predicts that the two components of Lucy's target Dinkinesh's secondary Salam \citep{lev23} should have evolved through BYORP rather than tides, which is necessary to have a non-asymptotic convergent evolution of two equal-mass satellites which presumably merged into Selam. Our model predicts that secondaries in wide asynchronous binaries have been stranded at large distances through BYORP, and then de-synchronized by YORP. This is similar to the model of \citep{jac14}, but with a negligible contribution from tides. Equal-sized pairs like Hermes \citep{pra16} may also hold the potential to resolve the degeneracy between tides and BYORP. Before either body's spin is synchronized evolution is purely through tides, while once they are doubly-synchronous the evolution is driven by BYORP only, offering us a potential way of separating the two effects.

\section{Conclusions}\label{sec:con}

Inspired by the data acquired by the DART spacecraft, as well as the ground-based observations before and after the DART impact, we propose two new hypotheses: 

1. Most secondaries in small binary asteroids may be ``debris piles'', meaning that they are rubble piles made exclusively of debris much smaller than the moon itself. ``Debris piles'' would be less blocky and closer to ellipsoids than previously observed rubble pile asteroids. Therefore, binary YORP (BYORP) effect in binary asteroids cannot be modelled using synthetic shapes based on single asteroids, which are less ellipsoidal. We propose that the surface roughness that gives rise to BYORP has a characteristic size on the order of 10-20 meters. Relative roughness is then inversely proportional to the satellite's size, making the BYORP effect even more size-dependent than previously thought. Combined with the recognition that BYORP coefficients are part of a distribution centered on the zero, this model can roughly explain the available observations of binary asteroid mutual orbital evolution.

2. We note that the post-impact Dimorphos appears to have a more energetically favored shape, appears to have kept synchronous rotation, and also exhibits significant changes in orbital period long after the impact. We conclude that a massive amount of dissipation has happened within Dimorphos after the impact, producing in a short time span the same consequences that are usually attributed to long-term tidal dissipation in binary asteroids. We propose that small binaries exhibit two modes regarding dissipation: most of the time binaries behave similarly to rigid bodies and are often dominated by BYORP, but occasional energetic events, such as impacts, can fluidize the system and lead to large-scale dissipation that can also include reshaping. The implication is that when observed binary asteroids have properties indicative of dissipation (such as synchronous rotation and low eccentricity), this may not be an indication of a long-term effect of tidal forces (as it often is for planetary satellites), but could be a result of a short-lived episode of high dissipation through material movement.

We hope that future observations and theoretical work will be able to support or contradict these hypotheses.

\begin{acknowledgments}
We thank two anomymous reviewers for their helpful comments that have improved the manuscript. M\'C is supported by NASA Solar System Workings program award 80NSSC21K0145. HFA was supported by the French government, through the UCA J.E.D.I. Investments in the Future project managed by the National Research Agency (ANR) with the reference number ANR-15-IDEX-01. KJW was supported in part through the NASA Solar System Exploration Research Virtual Institute node Project ESPRESSO, cooperative agreement number 80ARC0M0008. R.H.C. acknowledges that this material is based upon work supported by the National Science Foundation Graduate Research Fellowship Program under Grant No. DGE 2040434. Any opinions, findings, and conclusions or recommendations expressed in this material are those of the author(s) and do not necessarily reflect the views of the National Science Foundation. P.M. acknowledges funding support from CNES and ESA. 
\end{acknowledgments}

\bibliography{DART_refs}{}
\bibliographystyle{aasjournal}

\end{document}